\documentclass[twocolumn,epsfig,amsfonts,showpacs,prl]{revtex4}
\usepackage{epsfig}

\begin{document}

\title{Quantum corrections to the mass of self-dual vortices}

\author{A. Alonso Izquierdo$^{(1)}$,
W. Garcia Fuertes$^{(3)}$, M. de la Torre Mayado$^{(2)}$ and J.
Mateos Guilarte$^{(2)}$}

\affiliation{$^{(1)}$ Departamento de Matematica
Aplicada, Universidad de Salamanca, SPAIN \\
$^{(2)}$ Departamento de Fisica, Universidad de Salamanca, SPAIN
\\ $^{(3)}$ Departamento de Fisica, Universidad de Oviedo, SPAIN}

\begin{abstract}
The mass shift induced by one-loop quantum fluctuations on
self-dual ANO vortices is computed using heat kernel/generalized
zeta function regularization methods.
\end{abstract}


\maketitle

1. In this note we shall compute the one-loop mass shift for
Abrikosov-Nielsen-Olesen self-dual vortices in the Abelian Higgs
model. Non-vanishing quantum corrections to the mass of $N=2$
supersymmetric vortices were reported during the last year in
papers \cite{Vass} and \cite{Reb}. In the second paper, it was
found that the central charge of the $N=2$ SUSY algebra also
receives a non-vanishing one-loop correction which is exactly
equal to the one-loop mass shift; thus, one could talk about
one-loop BPS saturation. This latter result fits in a pattern
first conjectured in \cite{Reb1} and then proved in \cite{Shif}
for supersymmetric kinks. Recent work by the authors of the Stony
Brook/Viena group, \cite{Reb2}, unveils a similar kind of
behaviour of supersymmetric BPS monopoles in $N=2$ SUSY Yang-Mills
theory. In this reference, however, it is pointed out that
(2+1)-dimensional SUSY vortices behave not exactly in the same way
as their (1+1)- and (3+1)-dimensional cousins. One-loop
corrections in the vortex case are in no way related to an anomaly
in the conformal central charge, contrarily to the quantum
corrections for SUSY kinks and monopoles.

We shall focus, however, on the purely bosonic Abelian Higgs model
and rely on the heat kernel/generalized zeta function
regularization method that we developed in references \cite{Aai1},
\cite{Aai2} and \cite{Aai3} to compute the one-loop shift to kink
masses. Our approach profits from the high-temperature expansion
of the heat function, which is compatible with Dirichlet boundary
conditions in purely bosonic theories. In contrast, the
application of a similar regularization method to the
supersymmetric kink requires SUSY friendly boundary conditions,
see \cite{Vass1}. We shall also encounter more difficulties than
in the kink case due to the jump from one to two spatial
dimensions.

Defining non-dimensional space-time variables, $x^\mu \rightarrow
\frac{1}{ev} x^\mu$, and fields, $\phi \rightarrow
v\phi=v(\phi_1+i\phi_2)$, $A_\mu \rightarrow v A_\mu$, from the
vacuum expectation value of the Higgs field $v$ and the
$U(1)$-gauge coupling constant $e$, the action for the Abelian
Higgs model in (2+1)-dimensions reads:
\[
S= \frac{v}{e}\int d^3 x \left[ -\frac{1}{4} F_{\mu \nu} F^{\mu
\nu}+\frac{1}{2} (D_\mu \phi)^* D^\mu \phi - U(\phi,\phi^*)
\right]
\]
with $U(\phi,\phi^*)=\frac{\kappa}{8} (\phi^* \phi-1)^2$.
$\kappa=\frac{\lambda}{e^2}$ is the only classically relevant
parameter and measures the ratio between the masses of the Higgs
and vector particles; $\lambda$ is the Higgs field self-coupling.
For $\kappa=1$ one finds self-dual vortices with quantized
magnetic flux $g={2\pi l\over e}$, $l\in{\mathbb Z}$, and mass
$M_V=\pi |l|v^2$ as the solutions of the first-order equations
$D_1 \phi \pm i D_2 \phi=0$, $F_{12} \pm \frac{1}{2}
(\phi^*\phi-1) =0$, or,
\begin{eqnarray}
&&(\partial_1 \phi_1 +A_1 \phi_2)\mp(\partial_2 \phi_2-A_2
\phi_1)=0  \\
&&\pm(\partial_2 \phi_1 +A_2 \phi_2)+(\partial_1 \phi_2-A_1
\phi_1)=0 \\
&&F_{12}\pm\frac{1}{2}(\phi_1^2+\phi_2^2-1)=0
\end{eqnarray}
with appropriate boundary conditions: $\left. \phi^{*} \phi
\right|_{S_\infty} =1$, $D_i \phi |_{S_\infty} =(\partial_i \phi -
i A_i \phi)|_{S_\infty}=0$, that is, $\phi |_{S_\infty} = e^{il
\theta}$ and $A_i |_{S_\infty} =- i \phi^*\partial_i
\phi|_{S_\infty}$. In what follows, we shall focus on solutions
with positive $l$: i.e., we shall choose the upper signs in the
first-order equations.

2. ${\rm L}^2$-integrable second-order fluctuations around a
given vortex solution are still solutions of the first-order
equations with the same magnetic flux if they belong to the
kernel of the Dirac-like operator, ${\cal D}\xi (\vec{x})=0$,
\cite{Wein}
\[
{\cal D}=\left(\begin{array}{cccc} -\partial_2 & \partial_1 &
\psi_1 & \psi_2 \\ -\partial_1 & -\partial_2 & -\psi_2 & \psi_1
\\ \psi_1 & -\psi_2 & -\partial_2+V_1 & -\partial_1-V_2 \\ \psi_2 & \psi_1 &
\partial_1+V_2 & -\partial_2+V_1 \end{array}\right)
\]
where $\xi^T (\vec{x})= (a_1(\vec{x}),a_2(\vec{x}),
\varphi_1(\vec{x}),\varphi_2(\vec{x}))$. We denote the vortex
solution fields as $\psi=\psi_1+i\psi_2$ and $V_k$, $k=1,2$.
Assembling the small fluctuations around the solution $\phi
(\vec{x})=\psi (\vec{x})+\varphi (\vec{x})$,
$A_k(\vec{x})=V_k(\vec{x})+a_k(\vec{x})$ in a four column
$\xi(\vec{x})$,  the first component of ${\cal D}\xi$ gives the
deformation of the vortex equation (3), whereas the third and
fourth components are due  to the respective deformation of the
covariant holomorphy equations (2) and (1). The second component
sets the background gauge $B(a_k,\varphi;\psi)=\partial_k
a_k-(\psi_1\varphi_2-\psi_2\varphi_1)$ on the fluctuations. The
operators
\begin{widetext}
\begin{eqnarray*}
{\cal H}^+={\small \left(\begin{array}{cccc}
-\bigtriangleup +|\psi|^2 & 0 & -2\nabla_1\psi_2 & 2\nabla_1\psi_1
\\ 0 & -\bigtriangleup+|\psi|^2 & -2\nabla_2\psi_2 &
2\nabla_2\psi_1 \\ -2\nabla_1\psi_2 & -2\nabla_2\psi_2 &
-\bigtriangleup+{1\over 2}(3|\psi|^2+2V_kV_k-1) & -2V_k\partial_k
\\ 2\nabla_1\psi_1 & 2\nabla_2\psi_1 & 2V_k\partial_k  &
-\bigtriangleup+{1\over 2}(3|\psi|^2+2V_kV_k-1)
\end{array}\right)} \\
{\cal H}^-={\small \left(\begin{array}{cccc}
-\bigtriangleup+|\psi|^2 & 0 & 0 & 0 \\ 0 &
-\bigtriangleup+|\psi|^2 & 0 & 0  \\ 0 & 0 &
-\bigtriangleup+{1\over 2}(|\psi|^2+1)+V_kV_k & -2V_k\partial_k
\\ 0 & 0 & 2V_k\partial_k & -\bigtriangleup+{1\over
2}(|\psi|^2+1)+V_kV_k
\end{array}\right)} \qquad ,
\end{eqnarray*}
\end{widetext}
are defined as ${\cal H}^+={\cal D}^\dagger{\cal D}$ -the second
order fluctuation operator around the vortex in the background
gauge- and its partner ${\cal H}^-={\cal D}{\cal D}^\dagger$.

One easily checks that $\dim {\rm ker}{\cal D}^\dagger=0$. Thus,
the dimension of the moduli space of self-dual vortex solutions
with magnetic charge $l$ is the index of ${\cal D}$: ${\rm
ind}{\cal D}={\rm dim}{\rm ker}{\cal D}-{\rm dim}{\rm ker}{\cal
D}^\dagger$. We follow Weinberg \cite{Wein}, using the background
instead of the Coulomb gauge, to briefly determine ${\rm
ind}{\cal D}$. The spectra of the operators ${\cal H}^+$ and
${\cal H}^-$ only differ in the number of eigen-functions
belonging to their kernels. For topological vortices, we do not
expect pathologies due to asymmetries between the spectral
densities of ${\cal H}^+$ and ${\cal H}^-$ and thus ${\rm ind}\,
{\cal D}={\rm Tr}e^{-\beta{\cal H}^+}-{\rm Tr}e^{-\beta{\cal
H}^-}$. See \cite{Wein1,GM99} for the case of Chern-Simons-Higgs
topological vortices.

The heat traces ${\rm Tr}e^{-\beta{\cal H}^\pm}={\rm
tr}\int_{{\Bbb R}^2} \, d^2\vec{x} \, K_{{\cal
H}^\pm}(\vec{x},\vec{x};\beta)$ can be obtained from the kernels
of the heat equations:
\begin{eqnarray*}
&& \left(\frac{\partial}{\partial\beta}{\Bbb I}+{\cal H}^\pm
\right)K_{{\cal H}^\pm}(\vec{x},\vec{y};\beta )=0 \\
&& K_{{\cal H}^\pm}(\vec{x},\vec{y};0)={\Bbb I}\cdot
\delta^{(2)}(\vec{x}-\vec{y})
\end{eqnarray*}

Bearing in mind the structure ${\cal H}^\pm=-\bigtriangleup {\Bbb
I} +{\Bbb I}+Q^\pm_k(\vec{x})\partial_k+V^\pm(\vec{x})$, one
writes the heat kernels in the form:
\[
K_{{\cal
H}^\pm}(\vec{x},\vec{y};\beta)=C^\pm(\vec{x},\vec{y};\beta)K_{{\cal
H}_0}(\vec{x},\vec{y};\beta)
\]
with $C^\pm(\vec{x},\vec{x};0)={\Bbb I}$. $K_{{\cal
H}_0}(\vec{x},\vec{y};\beta)={e^{-\beta}\over
4\pi\beta}\cdot{\mathbb I}\cdot e^{-\frac{|\vec{x}-\vec{y}|
}{4\beta}}$ is the heat kernel for the Klein-Gordon operator
${\cal H}_0=(-\bigtriangleup+1){\Bbb I}$, which is the
second-order fluctuation operator around the vacuum in the
Feynman-'t Hooft renormalizable gauge, the background gauge in
the vacuum sector. $C^\pm(\vec{x},\vec{y};\beta)$ solve the
transfer equations:
\begin{eqnarray}
\left\{ {\partial\over\partial\beta}{\Bbb
I}+{x_k-y_k\over\beta}(\partial_k{\Bbb I}-{1\over
2}Q_k^\pm)-\bigtriangleup{\Bbb I}+  \right.&& \nonumber \\ \left.
+Q_k^\pm\partial_k+V^\pm \right\}C^\pm(\vec{x},\vec{y};\beta)=0
&& \label{eq:tran}
\end{eqnarray}
The high-temperature expansions
$C^\pm(\vec{x},\vec{y};\beta)=\sum_{n=0}^\infty
c_n^\pm(\vec{x},\vec{y})\beta^n$, $c_0^\pm(\vec{x},\vec{x})={\Bbb
I}$, trade the PDE (\ref{eq:tran}) by the recurrence relations
\begin{eqnarray}
[n{\Bbb I}+(x_k-y_k)(\partial_k{\Bbb I}-{1\over
2}Q_k^\pm)]c_n^\pm(\vec{x},\vec{y}) =&& \nonumber
\\=[\bigtriangleup{\Bbb I}
-Q_k^\pm\partial_k-V^\pm]c_{n-1}^\pm(\vec{x},\vec{y}) &&
\label{eq:rec}
\end{eqnarray}
among the coefficients with $n\geq 1$. Because
\begin{eqnarray}
{\rm Tr}e^{-\beta{\cal H}^\pm}&=&{e^{-\beta}\over
4\pi\beta}\sum_{n=0}^\infty\sum_{a=1}^4\int \, d^2x
\,[c_n]_{aa}^\pm(\vec{x},\vec{x})\beta^n  = \nonumber \\
&=&{e^{-\beta}\over 4\pi\beta}\sum_{n=0}^\infty\beta^n\sum_{a=1}^4
[c_n]_{aa}^\pm({\cal H}^\pm)
\end{eqnarray}
and $c_1^\pm(\vec{x},\vec{x})=-V^\pm(\vec{x})$, we obtain in the
$\beta=0$ -infinite temperature- limit:
\[
{\rm ind}{\cal D}={1\over 4\pi}{\rm tr}\left\{c_1({\cal
H}^+)-c_1({\cal H}^-)\right\}={1\over \pi}\int d^2x
V_{12}(\vec{x})=2l
\]
the dimension of the self-dual vortex moduli space is $2l$.

3. Standard lore in the semi-classical quantization of solitons
tells us that the one-loop mass shift comes from the Casimir
energy plus the contribution of the mass renormalization
counter-terms: $\Delta M_V=\Delta M_V^C+\Delta M_V^R$. The vortex
Casimir energy with respect to the vacuum Casimir energy is given
formally by the formula:
\[
\Delta M_V^C={\hbar m\over 2}\left[{\rm STr}^*\left({\cal
H}^+\right)^{{1\over 2}}-{\rm STr}\left({\cal H}_0\right)^{{1\over
2}}\right] \qquad ,
\]
where $m=ev$ is the Higgs and vector boson mass at the critical
point $\kappa=1$. We choose a system of units where $c=1$, but
$\hbar$ has dimensions of length $\times$ mass. The \lq\lq super
traces" encode the ghost contribution to suppress the pure gauge
oscillations: ${\rm STr}^*\left({\cal H}^+\right)^{{1\over
2}}={\rm Tr}^* \left({\cal H}^+\right)^{{1\over 2}}-{\rm
Tr}\left({\cal H}^G\right)^{{1\over 2}}$ and ${\rm STr}\left({\cal
H}_0\right)^{{1\over 2}}={\rm Tr}\left({\cal H}_0\right)^{{1\over
2}}-{\rm Tr}\left({\cal H}^G_0 \right)$. The trace for the ghosts
operators is purely functional: i.e., ${\cal
H}^G=-\bigtriangleup+|\psi|^2$, ${\cal H}_0^G=-\bigtriangleup +1$
are ordinary -non-matricial- Schrodinger operators. The star
means that the $2n$ zero eigenvalues of ${\cal H}^+$ must be
subtracted because zero modes only enter at two-loop order.

In a minimal subtraction renormalization scheme, one adds the
counter-terms ${\cal L}_{c.t.}^S = \hbar m I \left[|\phi|^2-1
\right]$, ${\cal L}_{c.t.}^A = -{1 \over 2}\hbar m I A_\mu A^\mu$
with $I=\int {d^2 {\vec k}\over (2 \pi)^2} {1\over \sqrt{\vec k
\cdot \vec k +1}}$ to cancel the divergences up to the
one-loop-order that arises in the Higgs tadpole and two-point
function, and in the two-point functions of the Goldstone and
vector bosons. Finite renormalizations are adjusted in such a way
that the critical point $\kappa=1$ is reached at first-order in
the loop expansion. Therefore, the contribution of the mass
renormalization counter-terms to the vortex mass is:
\[
\Delta M_V^R=\Delta M_{c.t.}^S+\Delta M_{c.t.}^A=\hbar\, m\,I \,
\Sigma (\psi,V_k)
\]
where $\Sigma (\psi,V_k)=\int \, dx^2 \, [(1-|\psi|^2)-{1\over 2}
V_kV_k]$.

We regularize both $\Delta M_V^C$ and $\Delta M_V^R$ by means of
generalized zeta functions. From the spectral resolution of a
Fredholm operator ${\cal H}\xi_n=\lambda_n\xi_n$, one defines the
generalized zeta function as the series $\zeta_{\cal H}(s)=\sum_n
{1\over\lambda_n^s}$, which is a meromorphic function of the
complex variable $s$. We can then hope that, despite their
continuous spectra, our operators fits in this scheme, and write:
\begin{eqnarray*}
\Delta M_V^C (s)&=&\frac{\hbar\mu}{2}\left({\mu^2\over
m^2}\right)^s\left\{\left(\zeta_{{\cal H}^+}(s)-\zeta_{{\cal
H}_G^+}(s)\right)+ \right. \\ &&+\left. \left(\zeta_{{\cal
H}_0^G}(s)-\zeta_{{\cal
H}_0}(s)\right)\right\} \\
\Delta M_V^R(s)& = &{\hbar\over m L^2} \zeta_{{\cal H}_0} (s)
\Sigma (\psi,V_k)
\end{eqnarray*}
where $\zeta_{{\cal H}_0} (s) = {m^2 L^2 \over 4 \pi} {\Gamma
(s-1) \over \Gamma(s)}$ and $\mu$ is a parameter of inverse length
dimensions. Note that $\Delta M_V^C=\lim_{s\rightarrow
-\frac{1}{2}}\Delta M_V^C(s)$, $\Delta M_V^R=\lim_{s\rightarrow
\frac{1}{2}}\Delta M_V^R(s)$ and $I=\lim_{s\rightarrow{1\over
2}}{1\over 2m^2L^2}\zeta_{{\cal H}_0}(s)$.

4. Together with the high-temperature expansion the Mellin
transform of the heat trace shows that
\[
\zeta_{{\cal H}}(s)={1\over\Gamma(s)}\sum_{n=0}^\infty \int_0^1 \,
d\beta \, \beta^{s+n-2}c_n({\cal
H})e^{-\beta}+{1\over\Gamma(s)}B_{{\cal H}}(s)
\]
is the sum of meromorphic and entire -$B_{\cal H}(s)$- functions
of $s$. Neglecting the entire parts and keeping a finite number of
terms $N_0$ in the asymptotic series for $\zeta_{\cal H}(s)$, we
find the following approximations for the generalized zeta
functions concerning our problem:
\begin{eqnarray*}
\zeta_{{\cal H}^+} (s) - \zeta_{{\cal H}_0} (s) & \simeq &
\sum_{n=1}^{N_0} \sum_{a=1}^4 [c_n]_{aa} ({\cal H}^+) \cdot
{\gamma[s+n-1,1] \over 4 \pi \Gamma(s)} \\
 \zeta_{{\cal H}_0^G} (s) -\zeta_{{\cal H}^G} (s)
& \simeq & - \sum_{n=1}^{N_0} c_n ({\cal H}^G) \cdot
{\gamma[s+n-1,1] \over 4 \pi \Gamma(s)} \quad ;
\end{eqnarray*}
$\gamma[s+n-1,1]=\int_0^1 \, d\beta \, \beta^{s+n-2}e^{-\beta} $
is the incomplete Gamma function, with a very well known
meromorphic structure. Contrarily to the (1+1)-dimensional case,
the value $s=-{1\over 2}$ for which we shall obtain the Casimir
energy  is not a pole.

Writing $\bar{c}_n=\sum_{a=1}^{4} [c_n]_{aa} ({\cal H}^+)-c_n
({\cal H}^G)$, the contribution of the first coefficient to the
Casimir energy
\[
\Delta M_V^{(1)C} (s)   \simeq {\hbar \over 2} \mu \left( {\mu^2
\over m^2}\right)^s \bar{c}_1 \cdot {\gamma[s,1/2] \over 4 \pi
\Gamma(s)}
\]
is finite at the $s\rightarrow -\frac{1}{2}$ limit
\[
\Delta M_V^{(1)C} (-1/2)  \simeq - {\hbar m \over 4 \pi } \Sigma
(\psi,V_k) \cdot {\gamma[-1/2,1] \over \Gamma(1/2)}
\]
and exactly cancels the contribution of the mass renormalization
counter-terms -also finite for $s={1\over 2}$-:
\begin{eqnarray*}
\Delta M_V^R (s)& \simeq &{\hbar m \over 4 \pi} \cdot \Sigma
(\psi,V_k) \cdot {\gamma[s-1,1]\over  \Gamma(s)} \\
\Delta M_V^R (1/2) & \simeq & {\hbar m \over 4 \pi} \cdot \Sigma
(\psi,V_k) \cdot
 {\gamma[-1/2,1]\over \Gamma(1/2)} \, .
\end{eqnarray*}
Subtracting the contribution of the $2l$ zero modes we finally
obtain the following formula for the vortex mass shift:
\begin{eqnarray}
 \Delta M_{V}& =&{\hbar m\over 2}\lim_{s\rightarrow
-\frac{1}{2}}\left[-2l\frac{\gamma[s,1]}{\Gamma(s)}+\sum_{n=2}^{N_0}
\bar{c}_n  {\gamma[s+n-1,1] \over 4
\pi \Gamma(s)}\right] \nonumber \\
&&\hspace{-1cm}= -{\hbar m \over 16 \pi^\frac{3}{2}}
\left[-2l\gamma[-{1\over 2},1]+ \sum_{n=2}^{N_0} \bar{c}_n
\gamma[n-3/2,1]\right] \label{eq:vorm}
\end{eqnarray}

5. Computation of the coefficients of the asymptotic expansion is
a difficult task; e.g. the second coefficient
\begin{eqnarray*}
&& c_2^+(\vec{x},\vec{x})= -{1\over 6}\bigtriangleup
V^+(\vec{x})+{1\over
12}Q_k^+(\vec{x})Q_k^+(\vec{x})V^+(\vec{x})-\\
&&- {1\over 6}\partial_kQ_k^+(\vec{x})V^+(\vec{x})+{1\over
6}Q_k^+(\vec{x})\partial_kV^+(\vec{x})+{1\over 2}[V^+]^2(\vec{x})
\end{eqnarray*}
Defining the partial derivatives of the coefficients at
$\vec{y}=\vec{x}$ as
\[
{}^{(\alpha_1,\alpha_2)}C_n^{ij}(\vec{x})=\lim_{\vec{y}\rightarrow
\vec{x}} \frac{\partial^{\alpha_1+\alpha_2}
[c_n]_{ij}(\vec{x},\vec{y})}{\partial x_1^{\alpha_1}\partial
x_2^{\alpha_2}}
\]
we write their recurrence relations
\begin{widetext}
\begin{eqnarray*} (k+\alpha_1&+&\alpha_2+1)
{}^{(\alpha_1,\alpha_2)}C_{k+1}^{ip}(\vec{x})=
{}^{(\alpha_1+2,\alpha_2)}C_{k}^{ip}(\vec{x})+
{}^{(\alpha_1,\alpha_2+2)}C_{k}^{ip}(\vec{x})- \\ &&-\sum_{j=1}^n
\sum_{r=0}^{\alpha_1}\sum_{t=0}^{\alpha_2} {\alpha_1 \choose r}
{\alpha_2 \choose t} \left[ \frac{\partial^{r+t}
Q^{ij}_1}{\partial x_1^r\partial x_2^t}
{}^{(\alpha_1-r+1,\alpha_2-t)}C_{k}^{jp}(\vec{x})+\frac{\partial^{r+t}
Q^{ij}_2}{\partial x_1^r\partial x_2^t}
{}^{(\alpha_1-r,\alpha_2-t+1)}C_{k}^{jp}(\vec{x})
\right]+\\
&&+\frac{1}{2}\sum_{j=1}^n
\sum_{r=0}^{\alpha_1-1}\sum_{t=0}^{\alpha_2} \alpha_1{\alpha_1-1
\choose r} {\alpha_2 \choose t}  \frac{\partial^{r+t}
Q^{ij}_1}{\partial x_1^r\partial x_2^t}
{}^{(\alpha_1-1-r,\alpha_2-t)}C_{k+1}^{jp}(\vec{x})+\\
&&+\frac{1}{2}\sum_{j=1}^n
\sum_{r=0}^{\alpha_2-1}\sum_{t=0}^{\alpha_1} \alpha_2{\alpha_2-1
\choose r} {\alpha_1 \choose t}  \frac{\partial^{r+t}
Q^{ij}_2}{\partial x_1^t\partial x_2^r}
{}^{(\alpha_1-t,\alpha_2-1-r)}C_{k+1}^{jp}(\vec{x})-\\&&-\sum_{j=1}^n
\sum_{r=0}^{\alpha_2}\sum_{t=0}^{\alpha_1}{\alpha_1 \choose
t}{\alpha_2\choose r}
 \frac{\partial^{r+t}
V^{ij}}{\partial x_1^t\partial x_2^r}
{}^{(\alpha_1-t,\alpha_2-r)}C_k^{jp}(\vec{x})
\end{eqnarray*}
\end{widetext}
starting from ${}^{(\beta,\gamma)}C_0^{jp}(\vec{x})$.

We notice that
$[{c}_n]_{jp}(\vec{x})={}^{(0,0)}C_n^{jp}(\vec{x})$ and thus
$[{c}_n]_{ii}({\cal H})=\int_{-\infty}^\infty d^2 x
[{c}_n]_{ii}(\vec{x})$.

Things are easier if we apply these formulae to cylindrically
symmetric vortices.  The ansatz $\phi(r,\theta) = f(r) e^{i
l\theta}$ and $rA_{\theta} (r,\theta) =l \alpha(r)$ plugged into
the first-order equations leads to:
\begin{equation}
{1\over r} {d \alpha \over d r}= \mp \frac{1}{2 l} (f^2-1) \quad
, \quad {d f\over d r}  = \pm \frac{l}{r} f(r)[1-\alpha(r)] \quad
. \label{eq:rrfo}
\end{equation}
Solutions of (\ref{eq:rrfo}) with the boundary conditions
${\displaystyle \lim_{r\rightarrow\infty}} f(r) = 1$,
${\displaystyle \lim_{r\rightarrow\infty}} \alpha(r) = 1$, zeroes
of the Higgs and vector fields at the origin, $f(0) =0$,
$\alpha(0)=0$, and integer magnetic flux, $eg= - \int_{r=\infty}
d\theta A_{\theta} = 2 \pi l$, can be found by a mixture of
analytical and numerical methods \cite{ViSh}. Henceforth, we shall
focus on the case $l=1$.

The heat kernel coefficients depend on successive derivatives of
the solution. This dependence can increase the error in the
estimation of these coefficients because we handle an
interpolating polynomial as the numerically generated solution,
and the derivation of such a polynomial introduces inaccuracies.
It is thus of crucial importance to use the first-order
differential equations (\ref{eq:rrfo}) in order to eliminate the
derivatives of the solution and write the coefficients as
expressions depending only on the fields. The recurrence formula
now gives the coefficients of the asymptotic expansion in terms
of $f(r)$ and $\alpha(r)$, e.g.:

{\footnotesize\begin{eqnarray*}&& \sum_{i=1}^4
[{c}_1]_{ii}(r,\theta)=5 -
\frac{2\,{\alpha(r)}^2}{r^2} - 5\,{f(r)}^2 \\
&& \sum_{i=1}^4 [{c_2}]_{ii}(r,\theta)=\frac{1}{12\,r^4}[37\,r^4
+ 4\,{\alpha(r)}^4 - 8\,r^2\,\left( -7 + 8\,r^2 \right) \,
{f(r)}^2 + \\&& +27\,r^4\,{f(r)}^4+8\,r^2\,\alpha(r)\,\left( 1 -
14\,{f(r)}^2 \right) + \\&& +8\,{\alpha(r)}^2\,\left( -2 - 3\,r^2
+ 9\,r^2\,{f(r)}^2 \right)]
\\&&\sum_{i=1}^4 [{c}_3]_{ii}(r,\theta)=
\frac{1}{120\,r^6}[ -4\,{\alpha(r)}^6 -
28\,r^2\,{\alpha(r)}^3\,\left( 2 + 5\,{f(r)}^2 \right) +\\ &&+
4\,{\alpha(r)}^4\,\left( 20 + 9\,r^2 + 32\,r^2\,{f(r)}^2 \right)
- 2\,r^2\,\alpha(r)\,\left( -4\,\left( 16 + 9\,r^2 \right)+\right.
\\&&\left. +\left(
32 + 331\,r^2 \right) \,{f(r)}^2 + 57\,r^2\,{f(r)}^4 \right)+
{\alpha(r)}^2\,\left( -256 - 144\,r^2 \right. \\ && \left.-
117\,r^4 +2\,r^2\,\left( 56 + 183\,r^2 \right) \,{f(r)}^2 +
99\,r^4\,{f(r)}^4 \right)  + r^4\,\left( -16 + \right.\\&& \left.+
151\,r^2 + \left( 392 - 321\,r^2 \right) \,{f(r)}^2 + \left( -20
+ 199\,r^2 \right) \,{f(r)}^4 \right. \\ && \left. -
29\,r^2\,{f(r)}^6 \right)] \qquad .
\end{eqnarray*}}
Plugging in these expressions the partially analytical partially
numerical solution for $f(r)$ and $\alpha(r)$, it is possible to
compute the coefficients -also for the ghost operator via similar
but simpler formulae- and integrate numerically them in the whole
plane. Thus, formula (\ref{eq:vorm})
\[
\frac{\Delta M_V}{\hbar m}=
\frac{-1}{16\pi^{\frac{3}{2}}}\sum_{n=2}^{N_0}\bar{c}_n
\gamma[-\frac{3}{2}+n,1] -\frac{1}{\sqrt{\pi}}
\]
provides us with the one-loop vortex mass shift, where we recall
that
\[
\bar{c}_n=\sum_{a=1}^{4} [c_n]_{aa} ({\cal H}^+)-c_n ({\cal H}^G)
\quad .
\]
The results are shown in the Table I:
\begin{table}[h]
  \caption{Seeley Coefficients and Mass Shift}
  \label{table:tabla1}
  \begin{ruledtabular}
    \begin{tabular}{ccc|cc}
$n$ & $\sum_{i=1}^4 {c}_n^{ii}({\cal H}^+)$ & $ {c}_n({\cal
H}^G)$ & $N_0$  & $\Delta M_V (N_0)/\hbar m$
\\ \hline
2 & 30.3513 & 2.677510 & 2 & -1.02814 \\
3 & 13.0289 & 0.270246 & 3 & -1.08241 \\
4 &  4.24732 &  0.024586 & 4 & -1.09191 \\
5 & 1.05946 &  0.001244 & 5 & -1.09350\\
6 &  0.207369 & 0.000013 & 6 & -1.09373
    \end{tabular}
  \end{ruledtabular}
\end{table}

\noindent The final value for the vortex mass at one-loop order
is:
\[
M_V=m\left(\frac{\pi v}{e}-1.09373 \hbar\right)+o(\hbar^2).
\]
The convergence up to the sixth order in the asymptotic expansion
is very good. We have no means, however, of estimating the error.
In the case of $\lambda (\phi)^4_2$ kinks we found agreement
between the result obtained by this method and the exact result up
to the fourth decimal figure, see \cite{Aai1} .


\end{document}